\documentclass[a4paper,11pt]{article}
\usepackage{jinstpub} % for details on the use of the package, please
\usepackage{lineno}

\usepackage{graphicx}%
\usepackage{multirow}%

\usepackage{amsmath,amssymb,amsfonts}%
\usepackage{amsthm}%
\usepackage{mathrsfs}%
\usepackage[title]{appendix}%
\usepackage{xcolor}%
\usepackage{textcomp}%
\usepackage{manyfoot}%
\usepackage{booktabs}%
\usepackage{algorithm}%
\usepackage{algorithmicx}%
\usepackage{algpseudocode}%
\usepackage{listings}%

\usepackage{subfigure,dcolumn}
\usepackage{epstopdf}
\usepackage{mhchem}
\usepackage{upgreek}
\usepackage{ulem}
\usepackage{url}

\usepackage{siunitx}
\newcommand{\kl}{K_{L}}
\newcommand{\EE}{e^+e^-}
\newcommand{\MM}{\mu^+\mu^-}

\newcommand{\bcntr}{\begin{center}}
\newcommand{\ecntr}{\end{center}}
\newcommand{\beq}{\begin{equation}}
\newcommand{\eeq}{\end{equation}}
\newcommand{\beqar}{\begin{eqnarray}}
\newcommand{\eeqar}{\end{eqnarray}}
\newcommand{\bitm}{\begin{itemize}}
\newcommand{\benu}{\begin{enumerate}}

\newcommand{\bitmb}{\begin{itemize}}
\newcommand{\benub}{\begin{enumerate}}
\newcommand{\eitm}{\end{itemize}}

\newcommand{\bfrm}{\begin{frame}}
\newcommand{\efrm}{\end{frame}}
\newcommand{\bct}{\begin{center}}
\newcommand{\ect}{\end{center}}
\newcommand{\bclm}{\begin{columns}}
\newcommand{\eclm}{\end{columns}}
\newcommand{\bpic}{\begin{overpic}}
\newcommand{\epic}{\end{overpic}}
\newcommand{\bblk}{\begin{block}}
\newcommand{\eblk}{\end{block}}

\newcommand{\eenu}{\end{enumerate}}

\newcommand{\ps}{\si{\pico\second}}
\newcommand{\s}{\si{\second }}

\newcommand{\m}{\si{\metre}}

\newcommand{\nm}{\si{\nano\metre}}
\newcommand{\um}{\si{\micro\metre}}
\newcommand{\mm}{\si{\milli\metre}}
\newcommand{\cm}{\si{\centi\metre}}

\newcommand{\V}{\si{\volt}}
\newcommand{\mV}{\si{\milli\volt}}

\newcommand{\ns}{\si{\nano\second}}

\newcommand{\ohm}{\si{\ohm}}

\newcommand{\hz}{\si{\Hz}}

\newcommand{\mhz}{\si{\mega\Hz}}

\newcommand{\gevcs}{\hbox{GeV}/c^2}
\newcommand{\gevc}{\hbox{GeV}/c}
\newcommand{\gev}{\hbox{GeV}}

\newcommand{\tot}{\rm TiO_2}
\newcommand{\pe}{p.e.}
\newcommand{\npe}{N_{\pe}}

\title{\boldmath Design and test for the CEPC muon subdetector based on extruded scintillator and SiPM}

\author[a]{Hongyu Zhang,}
\author[a]{Xiyang Wang,}
\author[a]{Weihu Ma,}
\author[a]{Shiming Zou,}
\author[a]{Deqing Fang,\note{Corresponding author.}}
\author[a,b]{Wanbing He,}
\author[a]{Xiaolong Wang,\note{Corresponding author.}}
\author[c,d,e]{Zhen Wang,}
\author[c,d,e]{Rui Yuan}
\author[f]{and Qibin Zheng}

\affiliation[a]{Key Laboratory of Nuclear Physics and Ion-beam Application (MOE) and Institute of Modern Physics, 
Fudan University, \\
220 Handan Road, Shanghai, 200433, China}

\affiliation[b]{Shanghai Research Center for Theoretical Nuclear Physics, NSFC and Fudan University, \\ 
2005, Songhu Rodad, Shanghai, 200438, China}

\affiliation[c]{Tsung-Tao Lee Institute, Shanghai Jiao Tong University, \\
520 Shengrong Road, Shanghai 201210, China}

\affiliation[d]{Institute of Nuclear and Particle Physics, School of Physics and Astronomy, \\
800 Dongchuan Road, Shanghai 200240, China}

\affiliation[e]{Key Laboratory for Particle Astrophysics and Cosmology (MOE), Shanghai Key Laboratory for Particle 
Physics and Cosmology (SKLPPC), Shanghai Jiao Tong University, \\
800 Dongchuan Road, Shanghai 200240, China}

\affiliation[f]{Laboratory of Radiation Detection and Medical Imaging and School of Health Science and Engineering, 
University of Shanghai for Science and Technology, \\
516 Jungong Rodad, Shanghai, 200093, China}

\emailAdd{xiaolong@fudan.edu.cn}

\abstract{
The integration of a scintillator, wavelength-shifting fiber, and silicon photomultiplier (SiPM) has demonstrated 
superior performance in the K-long and Muon detector (KLM) of the Belle II experiment. This study outlines our 
research and development (R\&D) initiatives aimed at harnessing similar detection technologies, incorporating a novel 
scintillator and SiPM, for potential use in a muon detector for the proposed Circular Electron Positron Collider 
(CEPC) experiment. Our R\&D activities have been focused on evaluating the efficacy of a newly developed 
$150~\cm$-long scintillator, alongside the NDL SiPM featuring a sensitive area of $3~\mm \times 3~\mm$, or the 
Hamamatsu MPPC with a $1.3~\mm \times 1.3~\mm$ sensitive surface. The project also includes the fabrication of a 
detector strip and the implementation of techniques designed to optimize light collection efficiency. Cosmic ray 
testing has shown that both NDL SiPMs and MPPCs are capable of highly efficient photon collection, achieving 
efficiencies significantly exceeding 90\% when employing a threshold of 8 photoelectrons. Additionally, the time 
resolution for detecting events at the distant end of a scintillator strip has been measured to be better than 
$1.7~\ns$. The remarkable performance observed lays the foundation for advancing R\&D including prototype modules 
aiming for reference Technical Design Report of CEPC detector recently. }

\keywords{Muon spectrometers; Scintillators and scintillating fibres and light guides; Photon detectors for UV, 
visible and IR photons (solid-state) (PIN diodes, APDs, Si-PMTs, G-APDs, CCDs, EBCCDs, EMCCDs, CMOS imagers, etc); 
Performance of High Energy Physics Detectors}

\arxivnumber{2312.02553} % only if you have one

\begin{document}

\maketitle

%\flushbottom

\section{Introduction}

In the framework of the Standard Model of particle physics, the Higgs boson ($H$) is pivotal in shedding light on the 
origin of mass for quarks and leptons.  With the Higgs boson's mass approximately $m_H \approx 125~\gevcs$, it 
enables the conceptualization of a new $\EE$ collider. This collider would operate at a center-of-mass energy of 
$\sqrt{s} \sim 240~\gev$, effectively serving as a Higgs factory. Several initiatives, including the International 
Linear Collider~\cite{ILC1,ILC2}, the Circular Electron Positron Collider (CEPC)~\cite{cepc1, cepc2}, and the Future 
Circular Collider~\cite{FCC}, have been proposed in pursuit of this goal. The CEPC proposal, initiated in 2013, has 
since been followed by extensive research and development (R\&D) efforts focused on both the accelerator and the 
detector components.

The primary process for Higgs boson generation via $\EE$ annihilation is $\EE\to H + Z^0$, allowing for the precise 
determination of the $H$ boson through the recoil of the $Z^0$ boson, which exhibits a branching fraction of 
$(3.3662\pm 0.0066)\%$ for decay into a $\MM$ pair~\cite{PDG}. Consequently, the muon detector is of paramount 
importance in a Higgs factory setting. Beyond the reconstruction of $Z^0 \to \MM$, the presence of muons in the final 
state is often regarded as the ``golden channel" in new particle searches~\cite{muonDet}. As outlined in the 
Conceptual Design Report for the CEPC, its muon detector is tasked with muon identification, calibrating shower 
leakage from calorimeters, and probing for long-lived particles~\cite{cepc2}. The muon detection system is expected 
to excel in detection efficiency, maintain a low rate of hadron misidentification, ensure precise position 
resolution, and achieve broad coverage. Additionally, clear muon signals are invaluable for enhancing the trigger 
system's performance. The proposed layout for the CEPC muon detector, consisting of a barrel and two endcaps, 
incorporates eight sensitive layers interleaved with iron absorbers. This configuration is designed to achieve high 
standalone muon detection efficiency (95\%), requiring minimum position resolutions $\sigma_{r\phi} = 2.0~\cm$, 
$\sigma_{z} = 1.5~\cm$, alongside precise time resolution and a rate capability ranging from $50-100~\hz/\cm^{2}$, 
thereby facilitating crucial muon momentum insights~\cite{cepc2}.

In this article, we detail the R\&D efforts for the CEPC muon detector, employing scintillator-based technology akin 
to the KLM of the Belle II experiment~\cite{Belle2_tdr}, which took reference to the T2K experiment~\cite{T2K,FGD}. During the initial Belle experiment, Resistive Plate 
Chambers (RPCs) were utilized within the KLM~\cite{Belle}. However, the prolonged recovery time of RPCs, when 
operating under high luminosity conditions, led to a noticeable drop in efficiency, prompting the partial 
substitution with scintillator modules. These modules, composed of plastic scintillator strips, wavelength-shifting 
(WLS) fibers, and silicon photomultipliers (SiPMs), have demonstrated effective performance in 
Belle II~\cite{klm_pid}. Their operation not only meets but also largely aligns with the stringent requirements set 
for a muon detector in the CEPC, showcasing their potential for future applications. Meanwhile, the technology of 
employing plastic scintillators with WLS fibers is well-established and extensively utilized, as exemplified by the 
Target Tracker detector in the long-baseline neutrino oscillation OPERA experiment~\cite{OPERA} or studies of photon 
collections with various setups~\cite{ScintAndFiber}. Compared to the OPERA design, which uses photomultiplier tubes 
(PMTs) for readout, utilizing SiPMs offers significant advantages, including a reduction in detector dead zones and 
the elimination of the high-voltage circuit required by PMTs. 

The CEPC muon detector will have a large volume while maintaining a low cost, thus emphasizing the use of 
domestically produced scintillators and SiPMs to reduce expenses. In collaboration with manufacturers in Beijing over 
the past few years, we have achieved outstanding performance from their products as presented in this article, 
laying the foundation for prototype modules aiming for reference Technical Design Report (RefTDR) of CEPC detector. 

\section{Belle II KLM and scheme of a detector channel}

\begin{figure}[htbp]
\bcntr
\includegraphics[height=4.5cm]{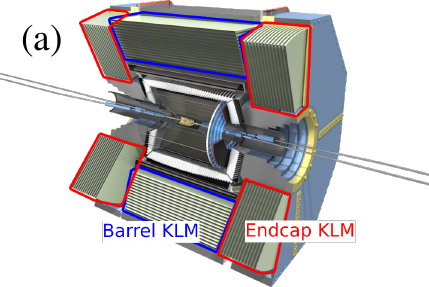}
\includegraphics[height=4.5cm]{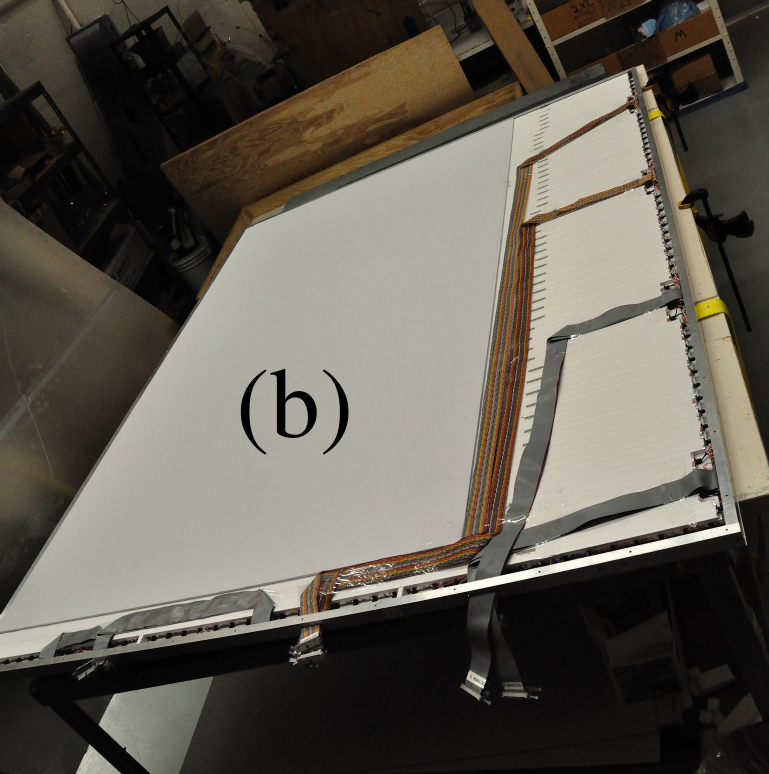}
\ecntr
\caption{The KLM system in the Belle II detector and a scintillator module for the barrel KLM.} 
\label{Belle2KLM}
\end{figure}

As the outermost component of the Belle II detector, the KLM system plays a crucial role in identifying $\kl$ mesons 
and muons, capable of detecting particles with momenta up to $4.5~\gevc$. The KLM extends over radii from $200$ to 
$340~\cm$ in its octagonal barrel section and from $130$ to $340~\cm$ in the forward and backward endcaps, as 
illustrated in Fig.~\ref{Belle2KLM}(a). The barrel configuration includes 15 layers of detector elements interleaved 
with 14 layers of yoke iron, whereas each endcap is structured with 14 detector and 14 iron layers. These yoke layers 
fulfill a dual purpose: facilitating magnetic flux return for the solenoid and aiding in the formation of hadronic 
cascades for $\kl$ detection. Detector panels, approximately $3.1~\cm$ in thickness, are strategically placed within 
the gaps of steel plates, each $4.7~\cm$ thick, to enhance particle detection efficiency. The detector's architecture 
incorporates scintillator strips within the two innermost layers of the barrel and throughout all layers of the 
endcaps, whereas the remaining 13 barrel layers utilize the legacy RPCs from the original Belle experiment. This 
design allows for precise two-dimensional localization of particle trajectories through orthogonal hits on the 
scintillator strips. The KLM boasts around 38,000 readout channels, half of which are dedicated to the 
scintillator-based detections, showcasing the system's extensive coverage and high-resolution particle tracking 
capabilities.

In the design of scintillator modules for the Belle II KLM detector, the cross-sectional dimensions of the 
scintillator strips are tailored for their specific locations: $4~\cm \times 1~\cm$ for the barrel and $4~\cm \times 
0.75~\cm$ for the endcaps, achieving a spatial resolution of approximately $\sigma_{s} \approx 1.2~\cm$. Each strip 
incorporates a Kuraray Y11(200)MSJ WLS fiber~\cite{y11} to efficiently capture and guide photons towards a SiPM, also 
known as a Multi-Pixel Photon Counter (MPPC: Hamamatsu S10362-13-050C), which is directly attached to one end of the 
WLS fiber. The MPPC enhances the initial signal by an internal preamplification factor of about 10, after which the 
signals are transmitted through ribbon cables to the external readout electronics located on the magnet yoke. 
Figure~\ref{Belle2KLM}(b) illustrates the interior of a scintillator module developed at Virginia Tech for the 
barrel section of KLM in 2013. Further details on the development and testing of the scintillator modules for the 
Belle II endcap KLM are provided in Ref.~\cite{eklm}.

\begin{figure}[htbp]
\bcntr
\includegraphics[height=4.5cm]{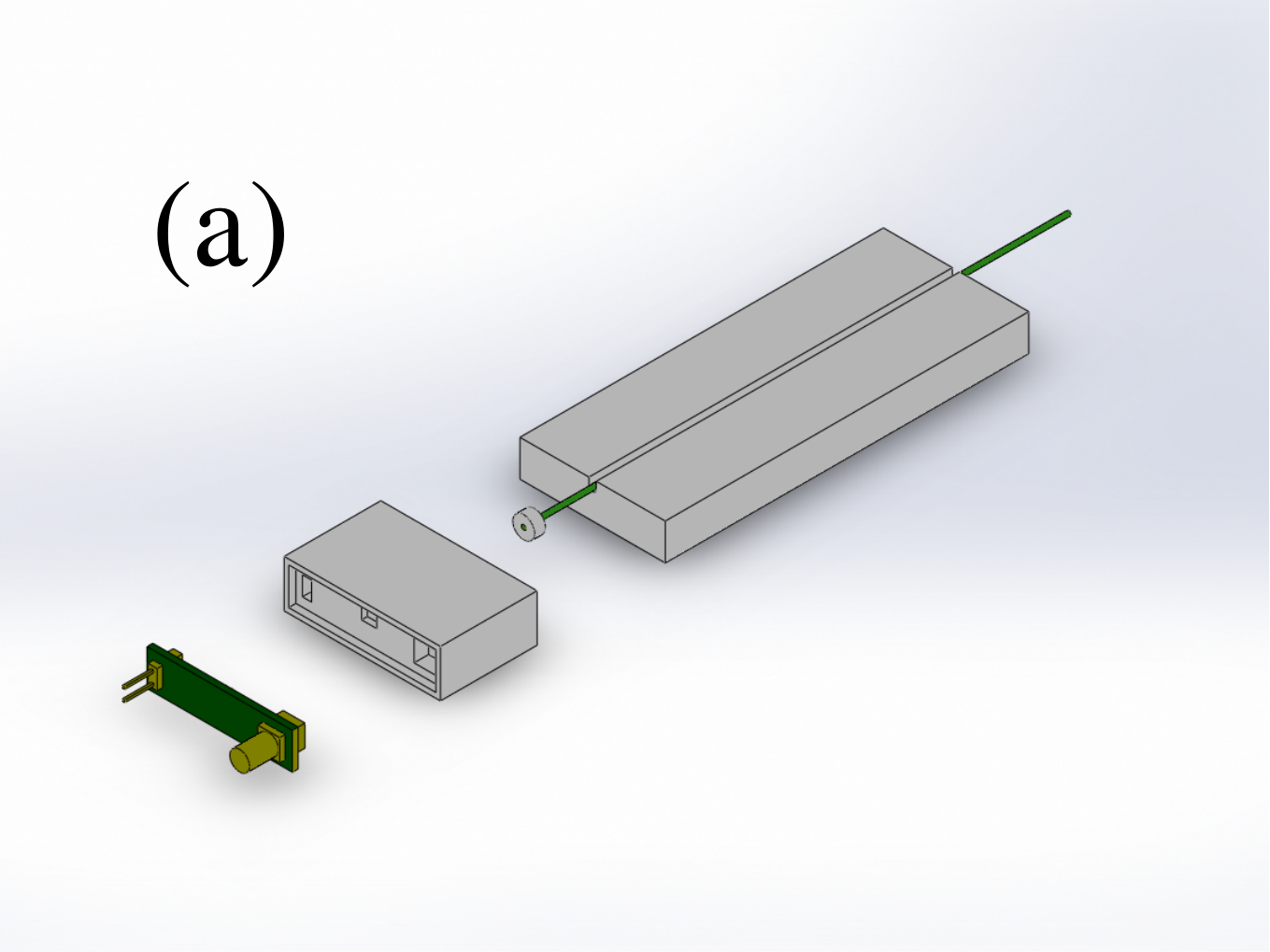} 
\includegraphics[height=4.5cm]{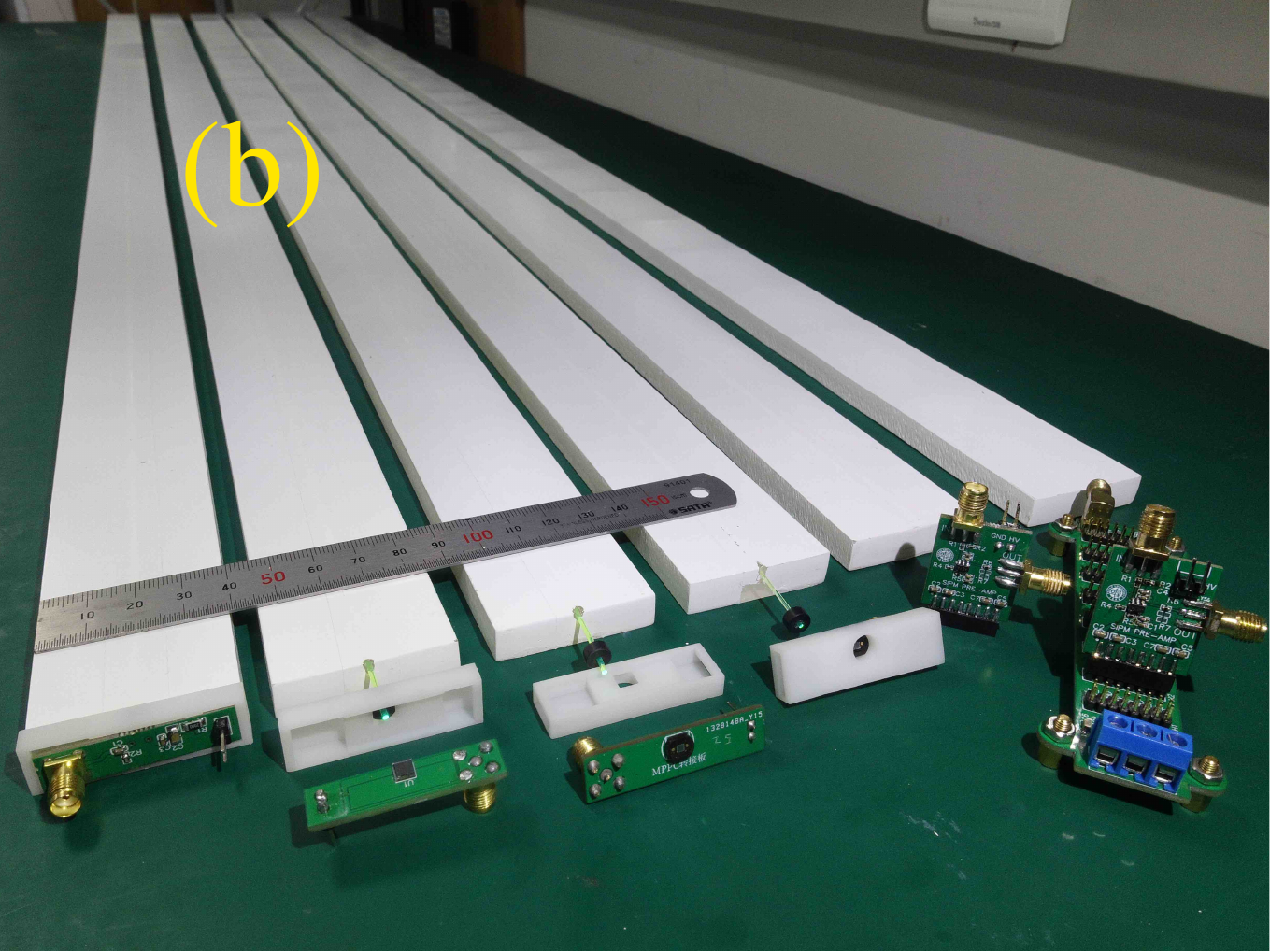} 
\ecntr
\caption{The architecture of a detector channel includes an elongated scintillator strip, measuring $150~\cm \times 
4~\cm \times 1~\cm$, equipped with an internal WLS fiber. A SiPM is positioned on a compact PCB featuring a 
preamplifier. Additionally, the design incorporates connectors for securing the fiber and facilitating its connection 
to the SiPM.} 
\label{structure}
\end{figure}

We adopted a design inspired by the Belle II KLM for our detector channel, utilizing an extruded scintillator as 
depicted in Fig.~\ref{structure}. This detector channel comprises a lengthy scintillator strip, measuring $1~\cm 
\times 4~\cm \times 150~\cm$, which houses a WLS fiber. A SiPM is mounted on a compact printed circuit board 
(PCB) and connected to the WLS fiber via a custom-designed coupling component, ensuring efficient signal transfer.

\section{The major components for a detector channel}

A detector channel comprises four key elements: a scintillator strip, a WLS fiber, a SiPM, and a front-end electronic 
readout equipped with a preamplifier. The novel readout design~\cite{pream} is compact, while the 
scintillator strips and SiPMs are custom-produced by manufacturers in Beijing. Additionally, we procured the 
identical WLS fiber utilized in the Belle II KLM experiment directly from Kuraray.

\subsection{Scintillator and WLS fiber}

Efficiencies exceeding 90\% have been achieved at the distant end with scintillator strips from established producers 
such as Fermilab and Uniplast for the Belle II KLM~\cite{Belle2_tdr}. In this R\&D work, the scintillator strips are 
produced by the GaoNengKeDi company~\cite{gnkd} using a polystyrene extrusion technique~\cite{extr}. The inherent 
short attenuation length of such scintillators, only several centimeters, poses challenges in long detector strips; 
this is circumvented by absorption of the scintillation light in a WLS fiber, which boasts an attenuation length of 
several meters. To accommodate the WLS fiber, a long groove, approximately $2~\mm$ wide and $4~\mm$ deep, is embedded 
within the scintillator. We evaluated both the BCF-92 fiber from Saint-Gobain~\cite{SGwls,WLS} and the Y11(200) fiber 
from Kuraray~\cite{y11}, with diameters of $1.0~\mm$ and $1.2~\mm$, respectively. Light yield performance of the two 
WLS fibers was assessed by comparing the activated pixels of the SiPM. The tests indicate that the light yield with 
Kuraray fiber is typically 2.3 times higher than that with Saint-Gobain fiber.

The Y11(200) fiber acts as a blue-to-green wavelength shifter with an emission spectrum peaking at 
$476~\nm$~\cite{y11}. According to its datasheet, Y11(200) fiber has an attenuation length exceeding $3.5~\m$. Light 
yields of the scintillator coupled with Y11(200) fiber were measured at various positions in cosmic ray tests and 
modeled with two exponential decay functions. The effective attenuation lengths derived from these functions are 
$(2.63\pm 0.37)~\m$ and $(5.8\pm 1.1)~\cm$, respectively.

\subsection{Hamamatsu MPPCs and NDL SiPMs}

A SiPM consists of a densely packed array of avalanche photodiodes, each operating in Geiger mode. This technology 
features exceptional photon detection efficiency, substantial gain, superior time resolution, low operational 
voltage, durability, and immunity to magnetic fields. The quantum efficiency of a SiPM typically approaches 50\%. 
Upon photon arrival, the SiPM's pixels trigger through the photoelectric effect, generating and amplifying electrons. 
This includes the generation of thermally induced electrons in the absence of photon input, known as dark counting. 
An electron thus generated and amplified in a SiPM, whether from an actual signal or dark counting, is termed a 
photo-electron ($\pe$). The output pulse amplitude is directly proportional to the number of photoelectrons ($\npe$), 
distinguishable in the SiPM's Analog-to-Digital Converter (ADC) signal distribution.

In our research, we utilize two variants of SiPMs: the S13360 series MPPC with a pixel size of $50~\um$, produced by 
Hamamatsu, and the EQR15 11-3030D-S series SiPM with a $15~\um$ pixel size, developed by the Novel Device Laboratory 
(NDL). The S13360 series MPPCs, in comparison to the S10362 series used in the existing KLM detector of Belle II, 
offer an identical sensitive area of $1.3~\mm \times 1.3~\mm$, but with lower operational voltages and decreased 
rates of cross-talk and dark counting (DCR). The EQR15 11-3030D-S series presents a larger sensitive area of 
$3.0~\mm \times 3.0~\mm$, significantly surpassing the cross-section of the Y11(200) WLS fiber.

Prior to their integration into our setup, we thoroughly investigate the Hamamatsu MPPCs and NDL SiPMs for key 
characteristics: breakdown voltage, gain, DCR, and cross-talk. Breakdown voltage refers to the reverse bias voltage 
at which the SiPM begins functioning via avalanche amplification. The gain of a SiPM quantifies the charge it 
accumulates from a single $\pe$ signal (SPE). Experimental evaluations demonstrate gains of $1.1 \times 10^5$ for the 
S13360 MPPC at an operational voltage of $57.0~\V$, and $0.7 \times 10^5$ for the EQR15 11-3030D-S at $27~\V$.

\begin{figure}[htbp]
\bcntr
\begin{minipage}[l]{0.49\columnwidth}
\centering
\includegraphics[height=4.2cm]{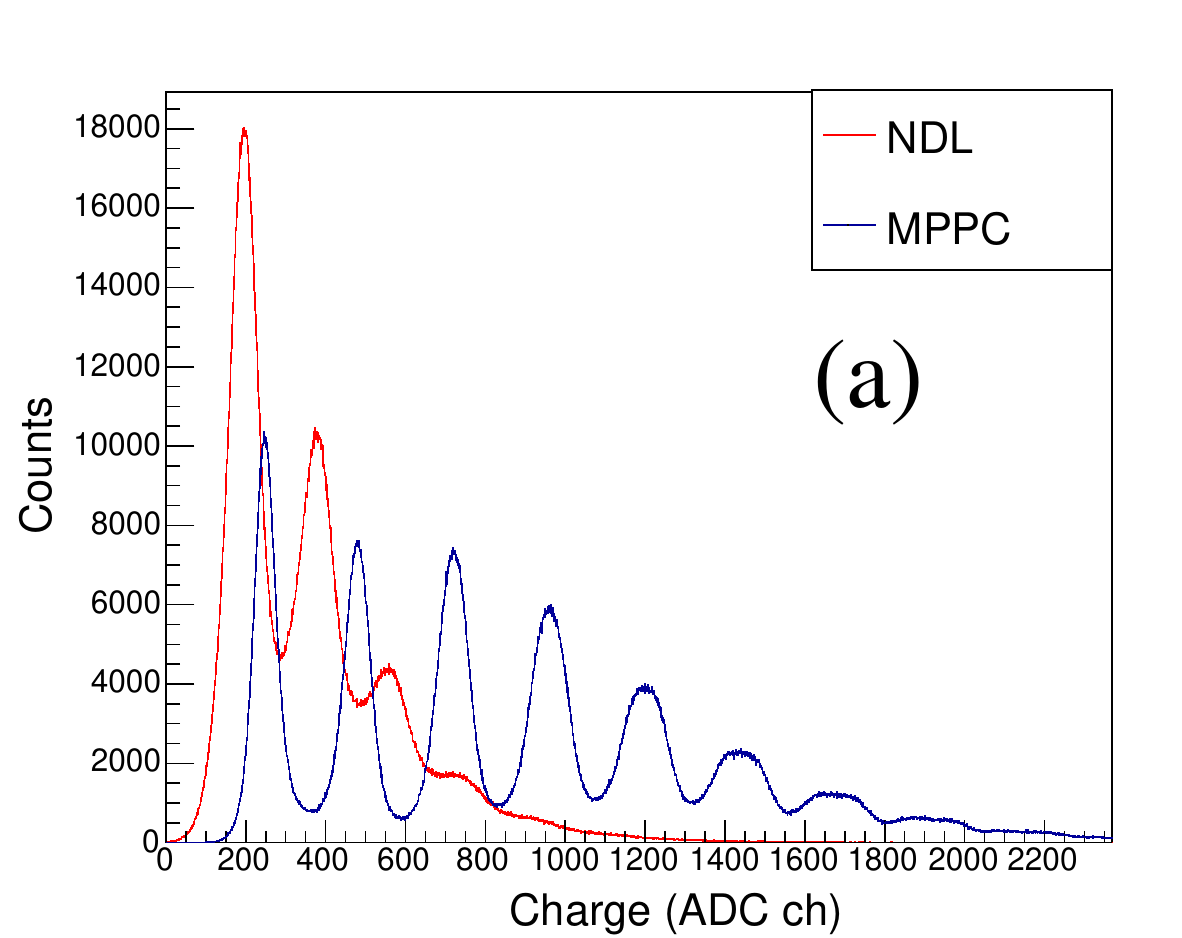}
\includegraphics[height=4.2cm]{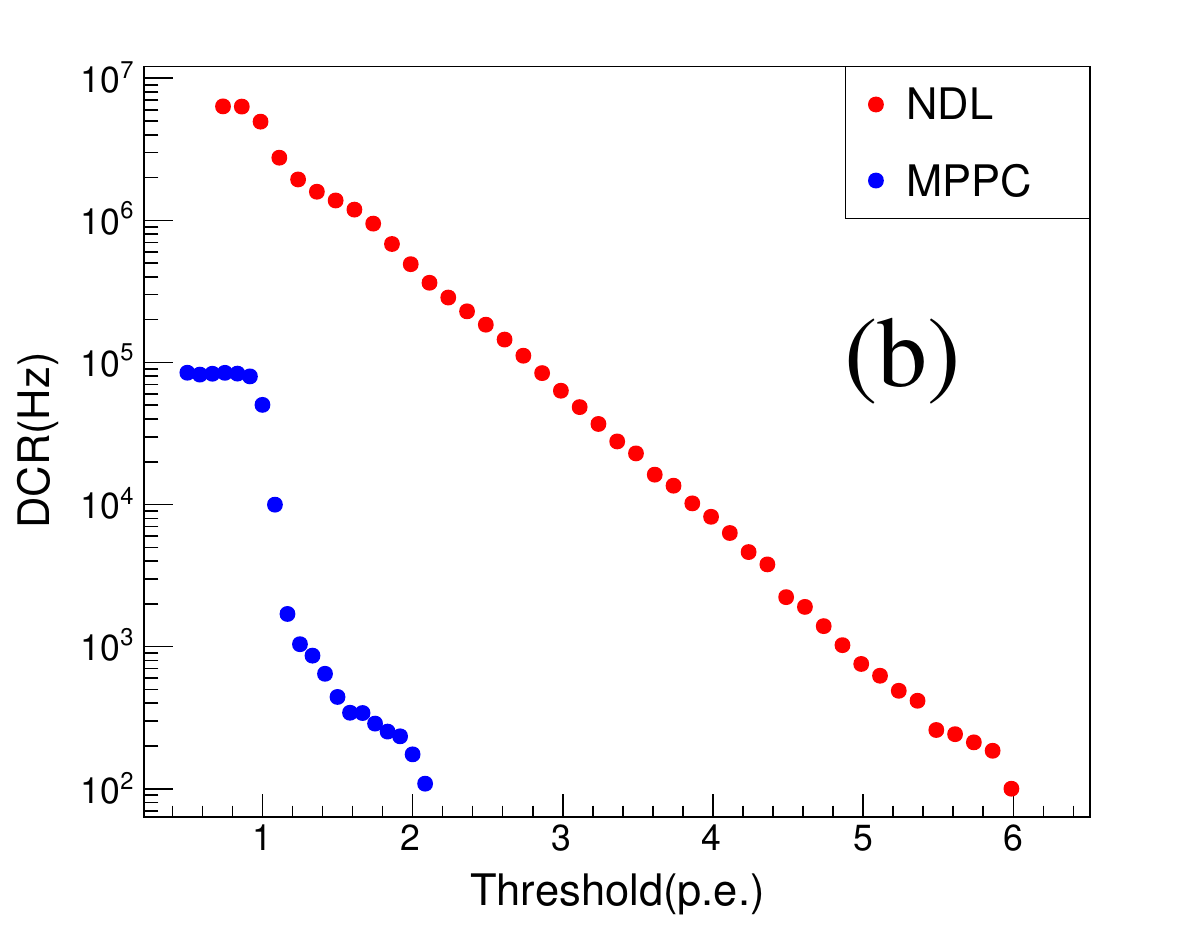} 
\end{minipage}
\begin{minipage}[r]{0.49\columnwidth}
\centering
\includegraphics[height=9.0cm]{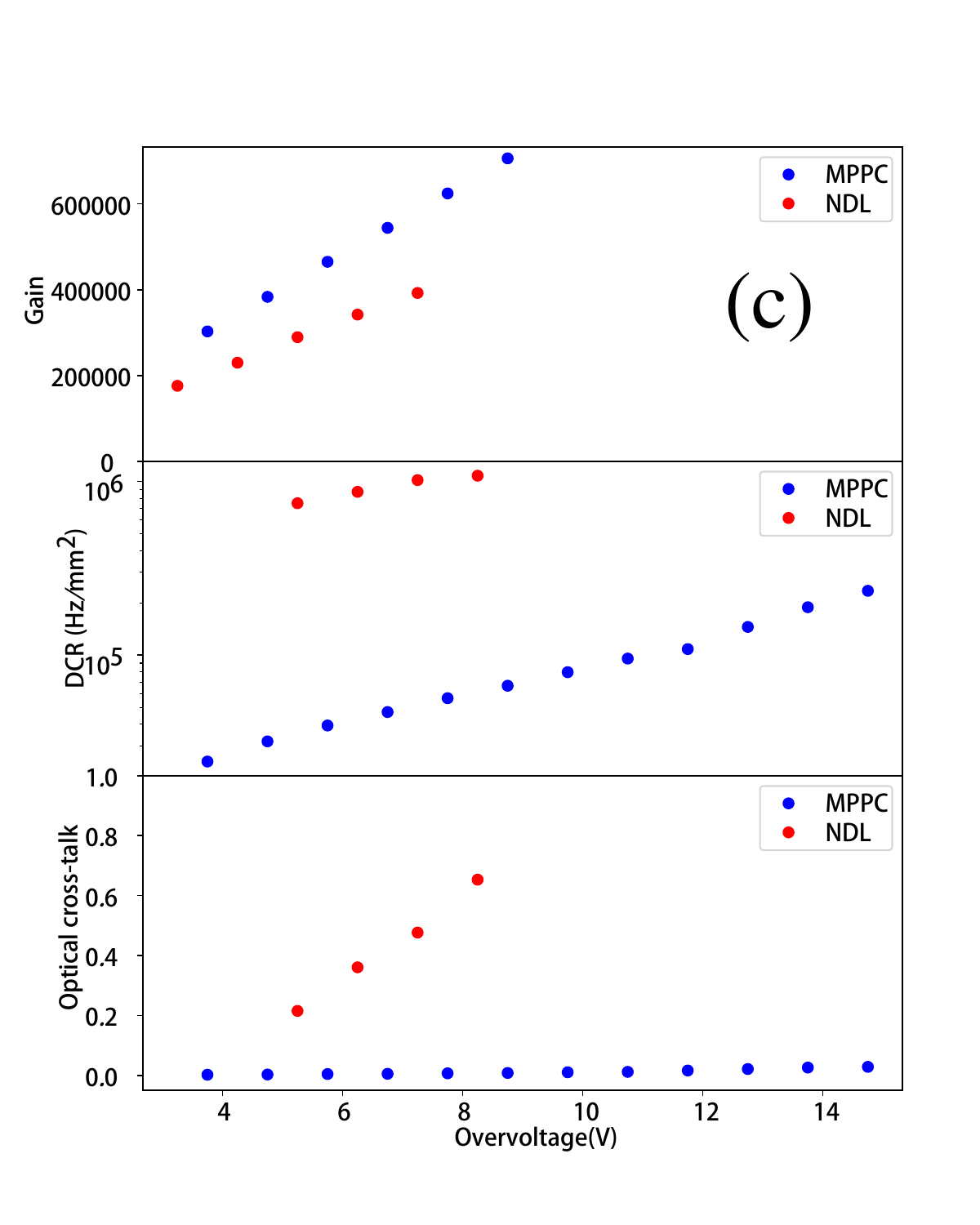}
\end{minipage}
\ecntr
\caption{Characteristics of the SiPMs employed in the R\&D for the CEPC Muon Detector. Plot (a) illustrates the 
photon detection spectra for the Hamamatsu MPPC at a bias of $56.0~\V$ and the NDL SiPM at $27.0~\V$. In these 
spectra, up to $8~\pe$ peaks are discernible for the MPPC, whereas for the NDL SiPM, peaks up to $4~\pe$ are 
observable. Plot (b) delineates the DCR for both MPPC and NDL SiPM as a function of the $\npe$ threshold. Lastly, 
plot (c) presents the relationship between the overvoltage and the corresponding variations in gain, DCR, and optical 
cross-talk for both MPPC and NDL SiPM devices.} 
\label{SiPM_cha}
\end{figure}

Dark counting in SiPMs presents a challenge for photon detection, particularly for SPE detection. The DCR of a SiPM 
is determined by counting the number of dark pulses surpassing a specified $\npe$ threshold. In this context, we set 
the threshold at $0.5~\pe$. Optical cross-talk is assessed by comparing the DCR for pulses exceeding thresholds of 
$1.5~\pe$ and $0.5~\pe$. As illustrated in Fig.~\ref{SiPM_cha}, NDL SiPMs exhibit significantly higher DCR and 
greater optical cross-talk compared to Hamamatsu MPPCs, a discrepancy attributed in part to their smaller pixel size 
and the epitaxial quenching resistor (EQR) technology being used. Generally, the dark count of a SiPM is 
closely related to the number of its pixels; the larger the sensitive area and the smaller pixel size of the SiPM, 
the higher the dark count. The NDL pixel count is two orders of magnitude larger than that of the MPPC, resulting in 
a much higher DCR. With our suggestion, NDL has been trying new products with larger pixel size 
($20~\um$)~\cite{NDL}, resulting in a much higher gain close to Hamamatsu MPPCs and a bit lower DCR.

\subsection{Readout electronics and data acquisition system}

SiPM operation generally necessitates a signal amplification circuit and an external power supply. We have developed 
a new preamplifier that achieves a time resolution superior to $50~\ps$ and a gain of 21~\cite{pream}. To enhance 
gain stability, we implement negative feedback in its design. Additionally, we engineered a power supply 
motherboard equipped with a voltage regulator module. This module provides both high voltage for the SiPM and 
low voltage for the preamplifier.

For signal waveform data collection, we employ a Tektronix MSO58 oscilloscope, which is PC-controlled and equipped 
with trigger logic~\cite{daq}. This setup allows us to efficiently store waveform data. The data acquisition system  
operates at a bandwidth of $20~\mhz$, with a sampling rate of $3.125~{\rm GS}/\s$, and captures data within a 
$400~\ns$ window. Through offline analysis of these waveforms, we extract comprehensive data on ADC, TDC, baseline 
stability, timing accuracy, and more. Specifically, we examin pulse heights or ADC counts to estimate the $\npe$ 
and utilize the leading edge or constant fraction discrimination method to derive precise timing information.

\section{Improvements for the photon collection}

The light yield of scintillators produced by GNKD, with a length of $20~\cm$ coupled with WLS fiber and PMTs, was 
studied in 2018~\cite{ScintAndFiber}. Our collaboration with GNKD began around 2018, during which we identified that 
their plastic scintillators exhibited insufficient light yield and a suboptimal reflective layer effect in 
1.5-meter-long samples, leading to poor light collection and low efficiency. Over the subsequent years, we have 
engaged with GNKD to enhance the performance of their scintillator samples.

\begin{figure}[htbp]
\bcntr
\includegraphics[height=3.2cm]{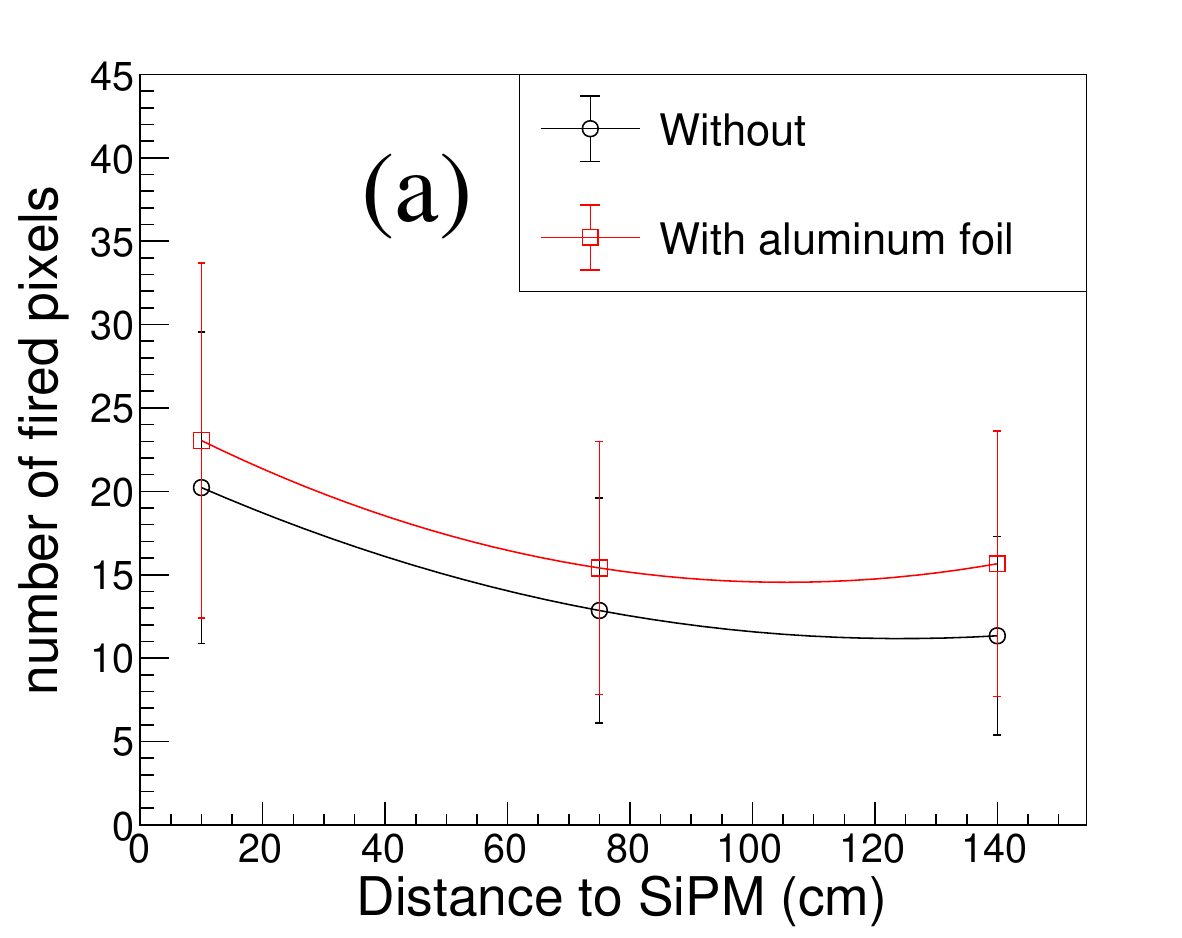} 
\includegraphics[height=3.2cm]{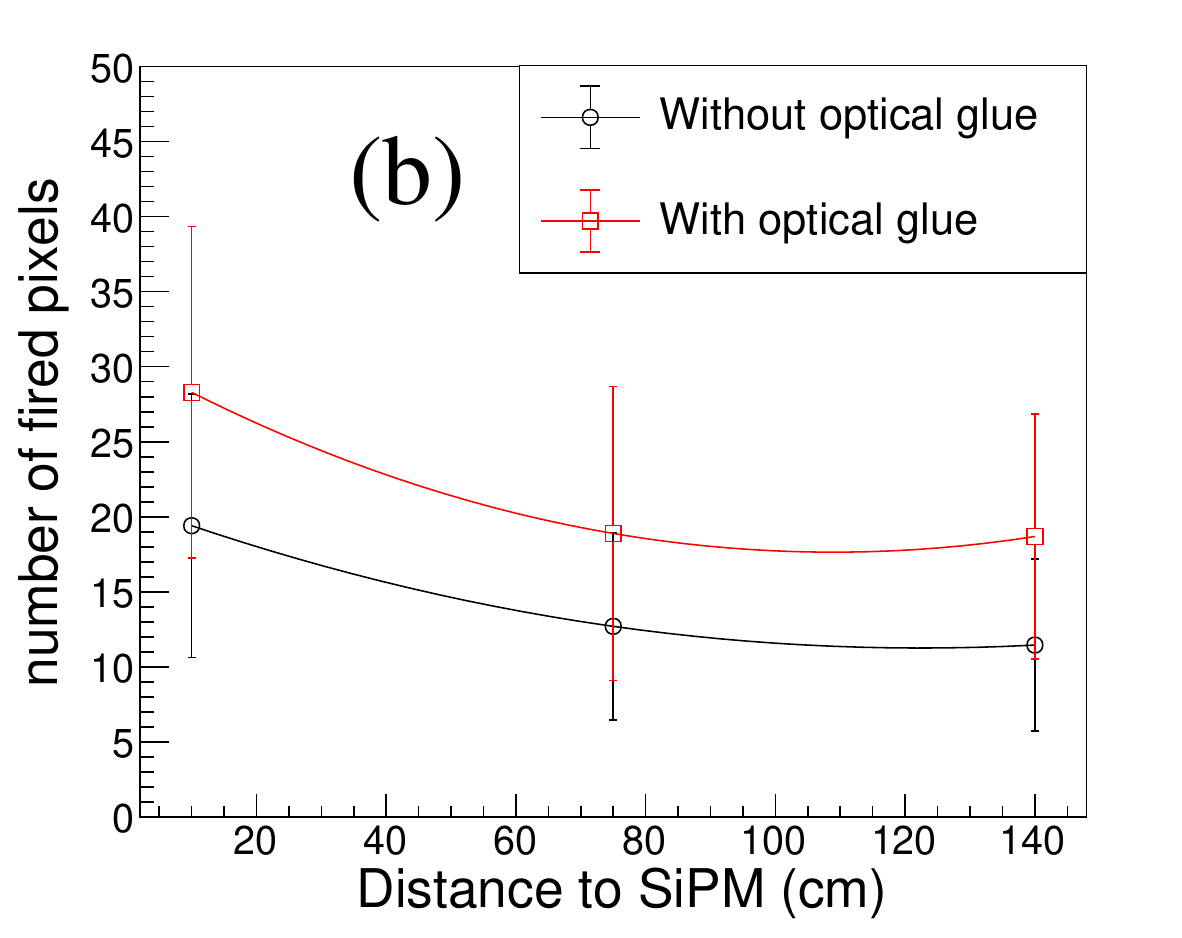} 
\includegraphics[height=3.2cm]{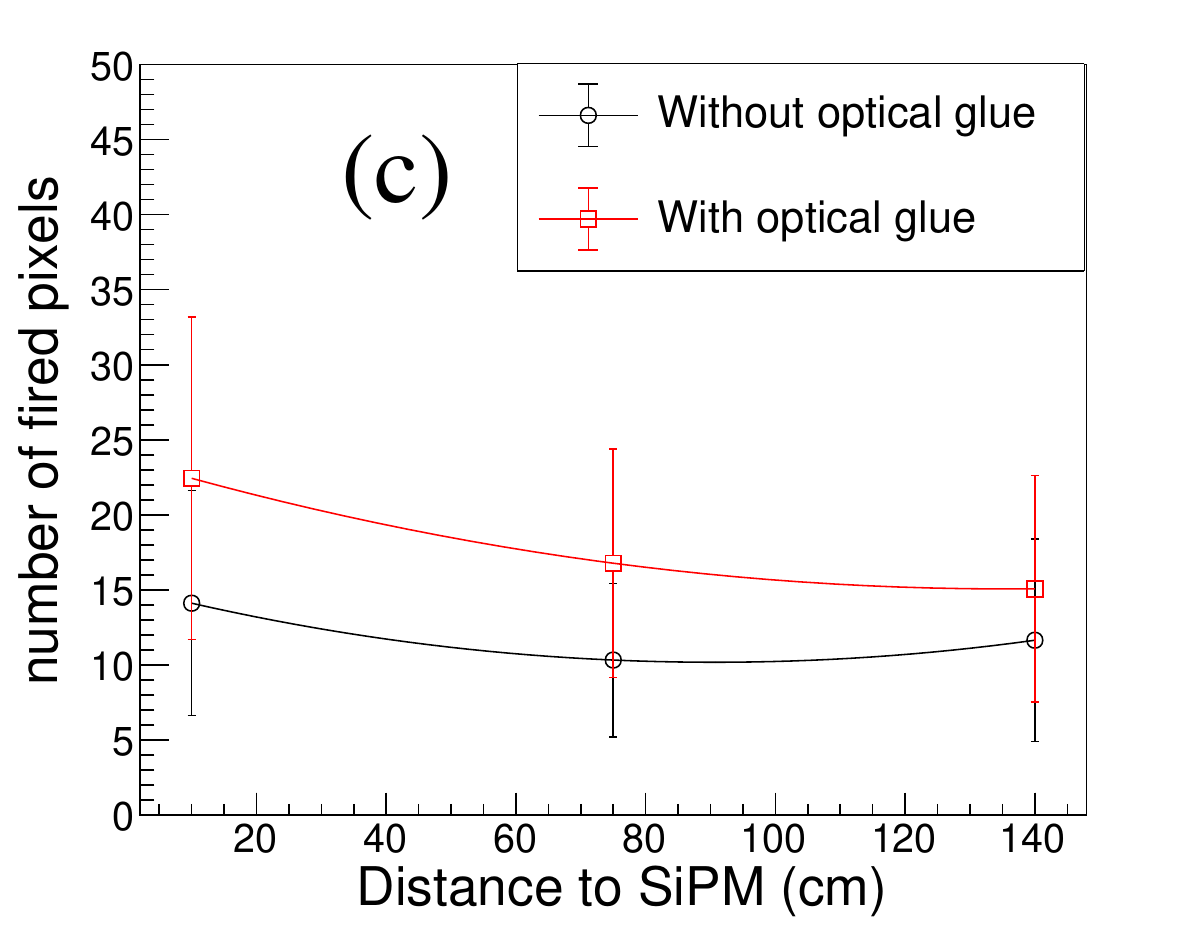} 
\ecntr
\caption{The distribution of the average $\npe$ varies according to the distance between the impact point on the 
scintillator strip and the SiPM. Plot (a) compares the scenarios with and without aluminum foil. Plots (b) and (c) 
showcase the effects of using optical glue with $\tot$ and Teflon, respectively. The data points and their error bars 
represent the mean values and standard deviations derived from fitting the $\npe$ distributions with Landau 
functions.}
\label{comparison}
\end{figure}

The coupling between the fiber, scintillator, and SiPM is crucial for the detector channel's performance, impacting 
detection efficiency and time resolution. We developed a coupling component to secure the WLS fiber and SiPM 
together, polished the end of the WLS fiber, applied a reflective layer to the scintillator in addition to the 
coating already done by the manufacturer, and filled the scintillator groove with optical glue. To evaluate 
performance, we triggered cosmic rays passing through the long strip at various positions using two short strips 
positioned above and below. The $\npe$ distribution from the SiPM signals was fitted with a Landau function. The 
plotted points and their error bars in Fig.~\ref{comparison} represent the fitted results, indicating the number of 
activated pixels. Here, the $y$-axis value and the error bar of each point represent the mean and the standard 
deviation ($\sigma$) of the Landau function.

The coupling mechanism used in the Belle II experiment has been enhanced in our R\&D to streamline the assembly of 
the detection system and enhance photon detection efficiency. As illustrated in Fig.~\ref{structure}, this modified 
component ensures a robust connection between the WLS fiber's end and the SiPM's sensitive area. These components are 
fabricated using a 3D printer for precision and consistency.

The fiber's end is polished with sandpaper, and we find that a 2000 mesh provides a sufficiently smooth surface. This 
refinement is particularly beneficial for employing the MPPC S13360, which has a sensitive area of $1.3~\mm \times 
1.3~\mm$. In the Belle II experiment, the proximity between the fiber and the SiPM was found to influence photon 
collection by approximately 37\%, primarily due to diffuse transmission at the fiber's end. The NDL EQR15 11-3030D-S, 
featuring a much larger sensitive area of $3.0~\mm \times 3.0~\mm$, demonstrates that meticulous polishing of the 
fiber's end is not critical for enhancing photon collection with this model.

Reflective coatings such as $\tot$ and Teflon are applied to the scintillator strips by GNKD to enhance light 
reflection. The coatings are done by the manufacturer in its workshop after the extruded scintillator strips 
being produced. Furthermore, we use aluminum foil and black plastic tapes to isolate the strips from ambient light 
and minimize cross-talk. Our comparisons indicate that aluminum foil not only shields against external light 
interference but also augments light collection, boosting $\npe$ by 38\% at the strip's far end, as depicted in 
Fig.~\ref{comparison}(a). Tests comparing reflective coatings shows that $\tot$ slightly outperforms Teflon, as 
presented in Figs.~\ref{comparison}(b) and \ref{comparison}(c). However, Teflon offers production simplicity and 
cost-effectiveness, suggesting that a thicker Teflon layer could be advantageous for further optimization. A 38\% 
improvement on the light collection means that there is potential to increase the quality of the coating at the 
manufacturer, so GNKD recently built a much larger new workshop focused on the application of spray-on reflective 
coatings, enabling the spraying of multiple layers of Teflon to enhance the reflectivity. 

The SYLGARD-184 optical glue~\cite{glue} is utilized to enhance the light-transfer coupling between the fiber and the 
scintillator. The commercially available SYLGARD-184 has a refractive index of approximately 1.4, which is identical 
to that of the extruded scintillator. In tests with $\tot$-coated strips, using optical glue resulted in a 
significant 63\% increase in $\npe$ at the strip's far end, illustrated in Fig.~\ref{comparison}(b). Similarly, 
Teflon-coated strips showed a 30\% improvement, as seen in Fig.~\ref{comparison}(c), underscoring the efficacy of 
optical glue in enhancing photon collection efficiency.

\section{Performance of an array of scintillator strips in cosmic ray tests}

\begin{figure}[htbp]
\bcntr
\includegraphics[height=4.5cm]{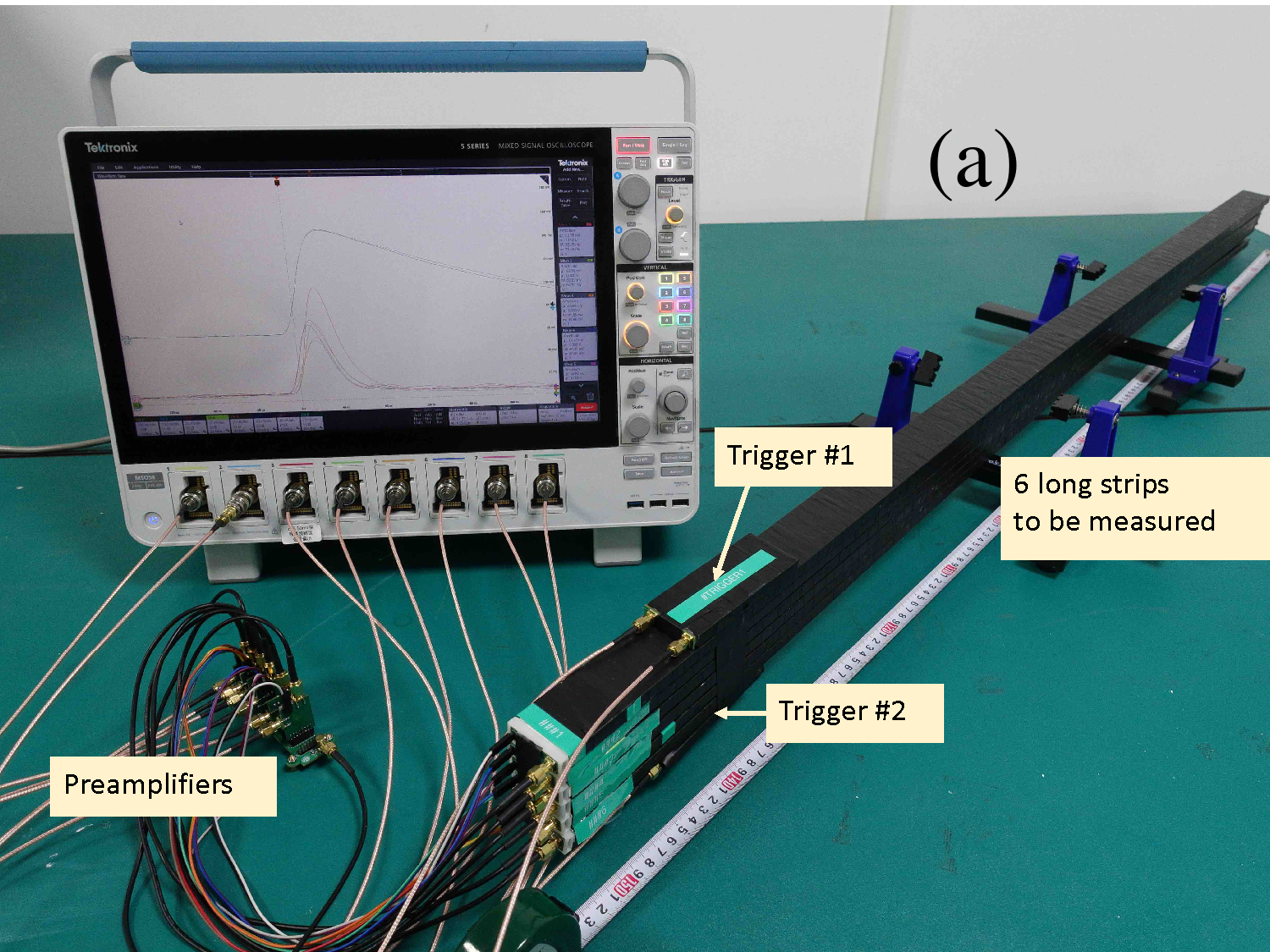} 
\includegraphics[height=4.5cm]{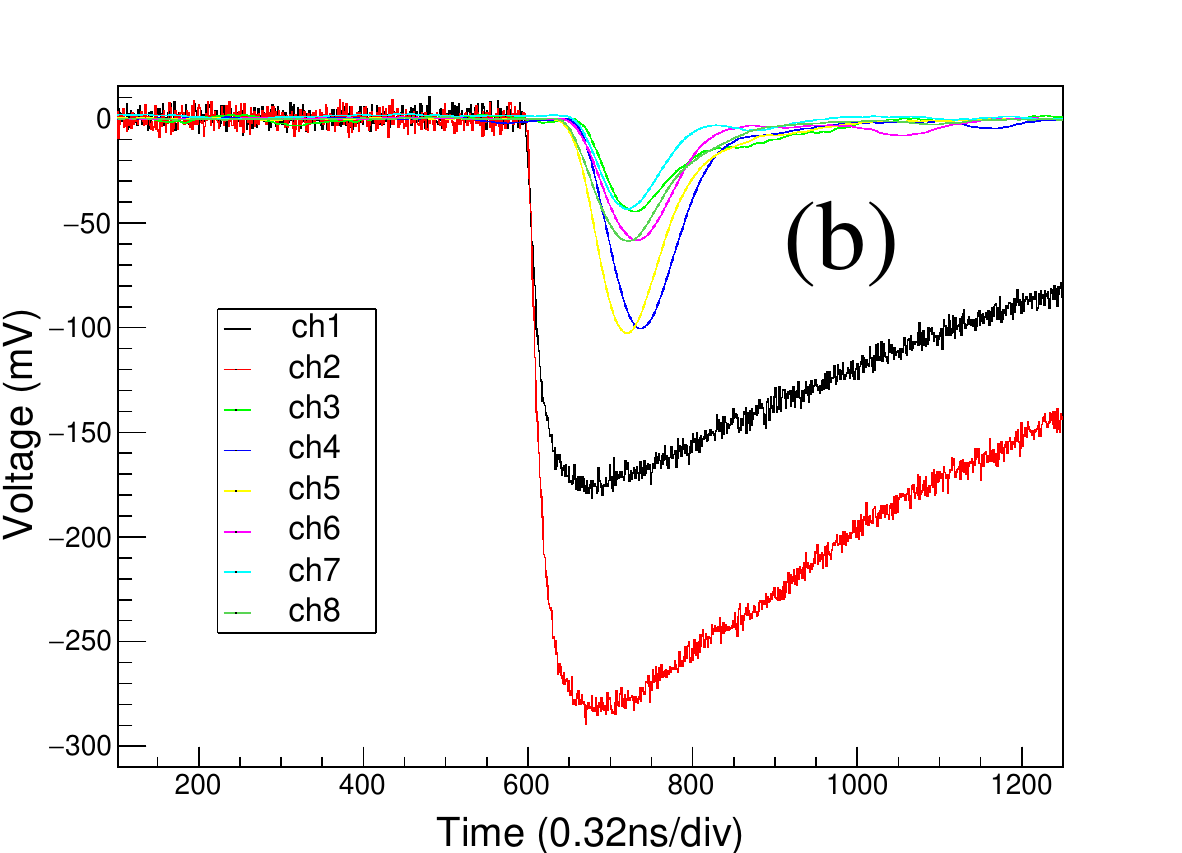} 
\ecntr
\caption{Setup for the cosmic ray test of six long strips, incorporating preamplifiers and an oscilloscope, alongside 
the oscilloscope-captured waveforms of six channels and two trigger channels at the distal end. In plot (b), ch1 and 
ch2 correspond to Trigger\#1 and Trigger\#2, as depicted in plot (a), with ch3-ch8 representing the six long strips. 
The pulse profiles for both NDL and MPPC display resemblance.}
\label{6chset}
\end{figure}

Similar to the KLM in the Belle II experiment as depicted in Fig.~\ref{Belle2KLM}, a muon detector usually comprises 
multiple layers of detector modules for effective charged track identification. Drawing from our experience with the 
KLM, the cornerstone of muon detector R\&D, particularly when utilizing extruded scintillators, is achieving high 
efficiency across all scintillator strips. To evaluate collective performance, we assembled an array of six detector 
channels and conducted cosmic ray tests, as illustrated in Fig.~\ref{6chset}(a). For triggering, we positioned two 
short strips, each $10~\cm$ in length, vertically at both ends of the six longer strips. The optimal operating 
voltages determined for the MPPC S13360 and NDL EQR15 11-3030D-S SiPMs are $56~\V$ and $27~\V$ respectively, based on 
their performance metrics presented in Fig.~\ref{SiPM_cha}. The signal amplitudes for SPE events from these SiPMs 
approximate $2~\mV$. Figure~\ref{6chset}(b) displays typical signal waveforms captured from the two trigger strips 
and the six detector strips.

\begin{figure}[htbp]
\bcntr
\includegraphics[width=0.3\textwidth]{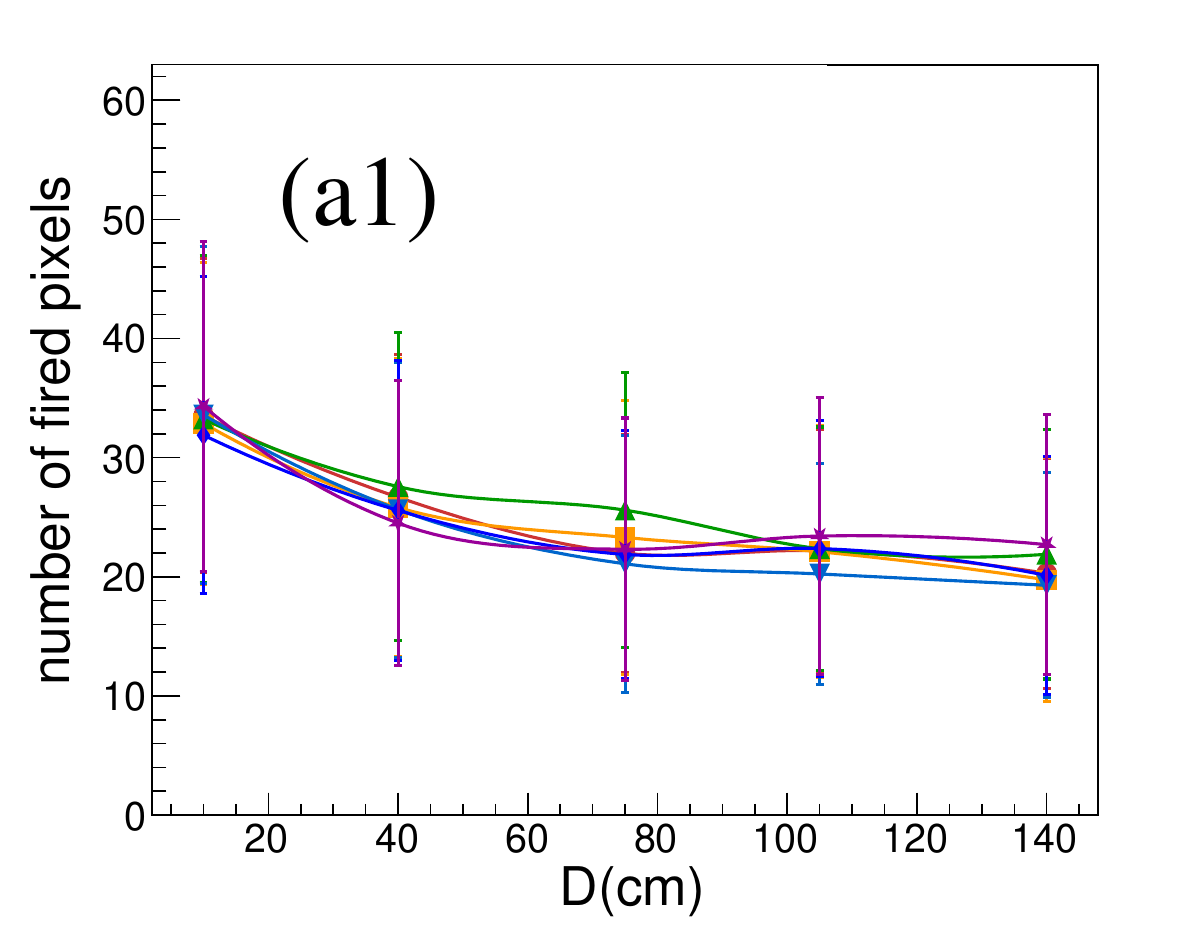} 
\includegraphics[width=0.3\textwidth]{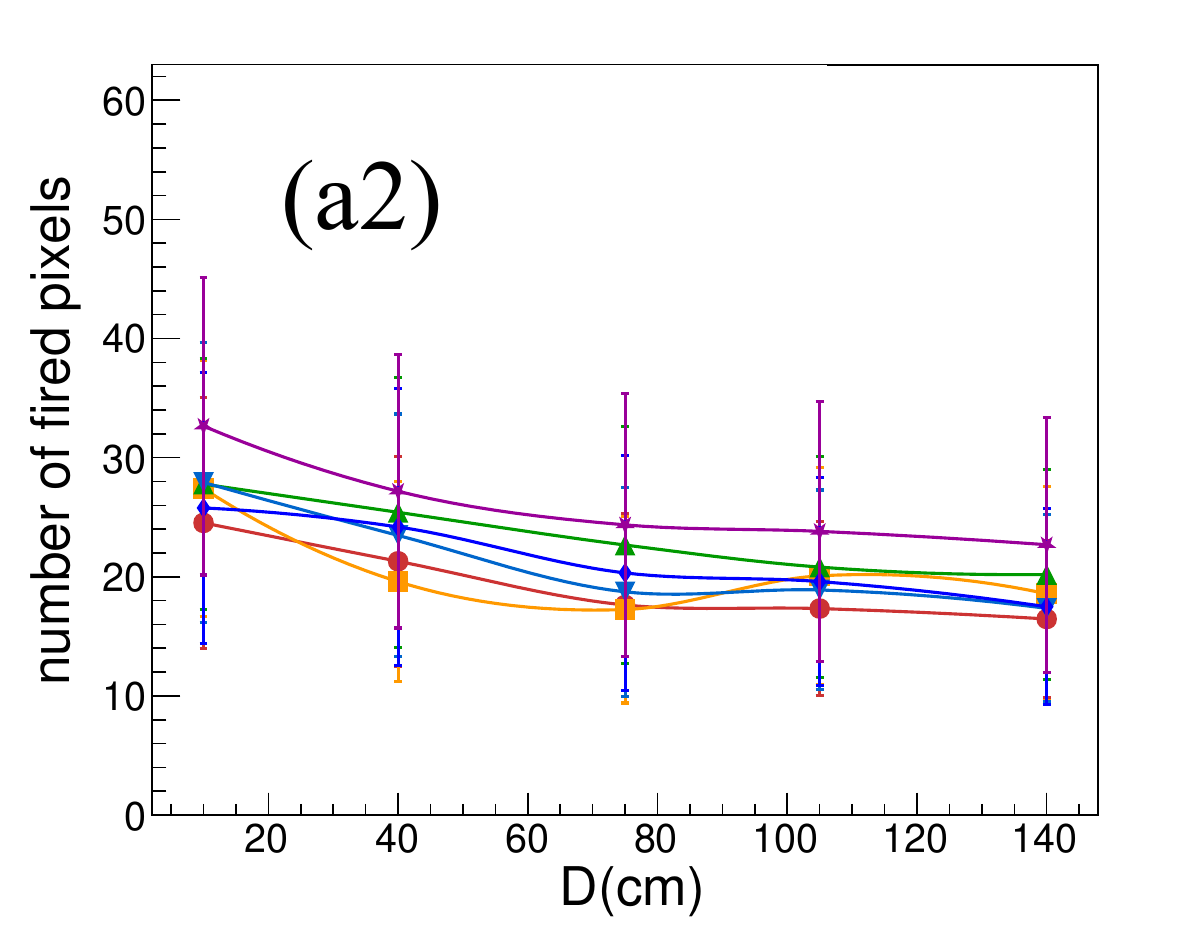} \\
\includegraphics[width=0.3\textwidth]{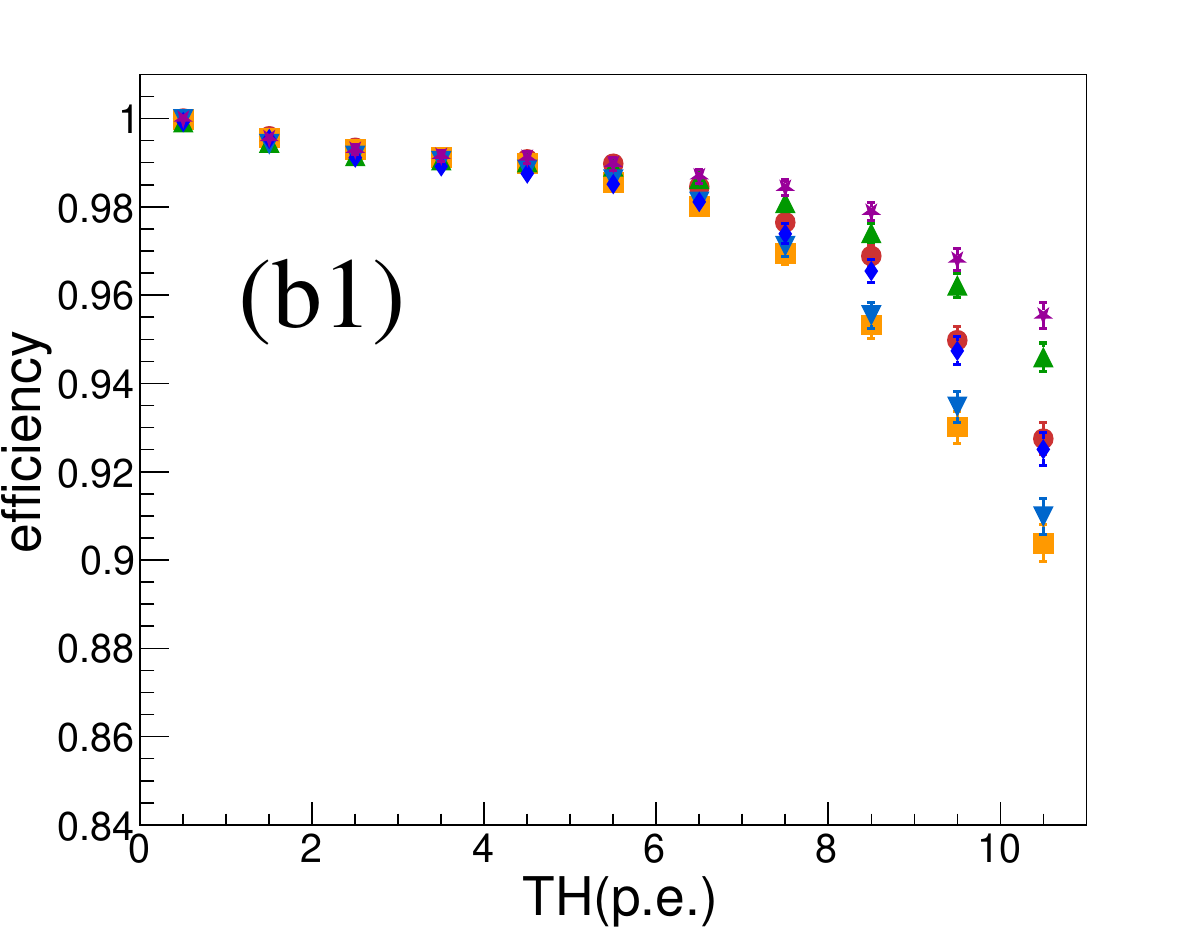} 
\includegraphics[width=0.3\textwidth]{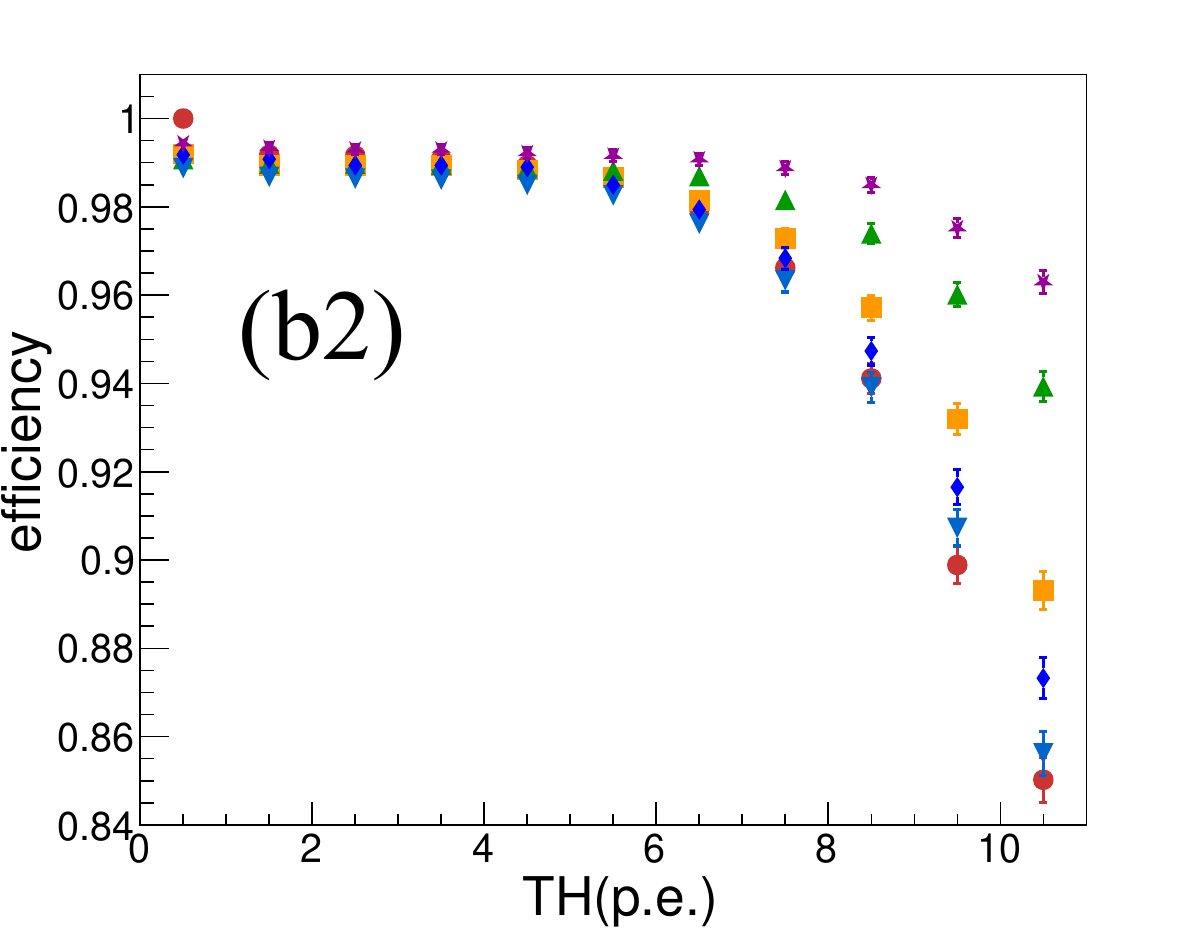} \\
\includegraphics[width=0.3\textwidth]{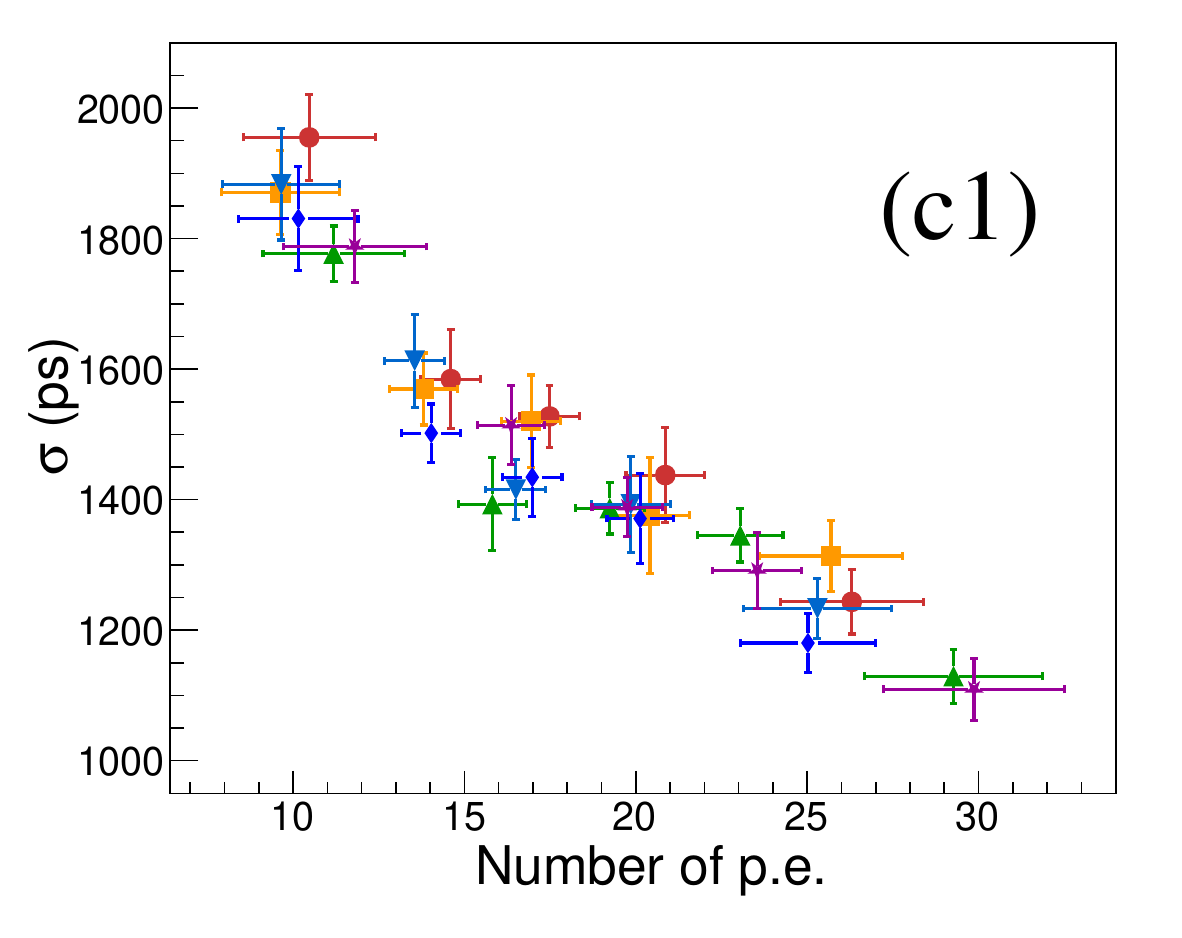}
\includegraphics[width=0.3\textwidth]{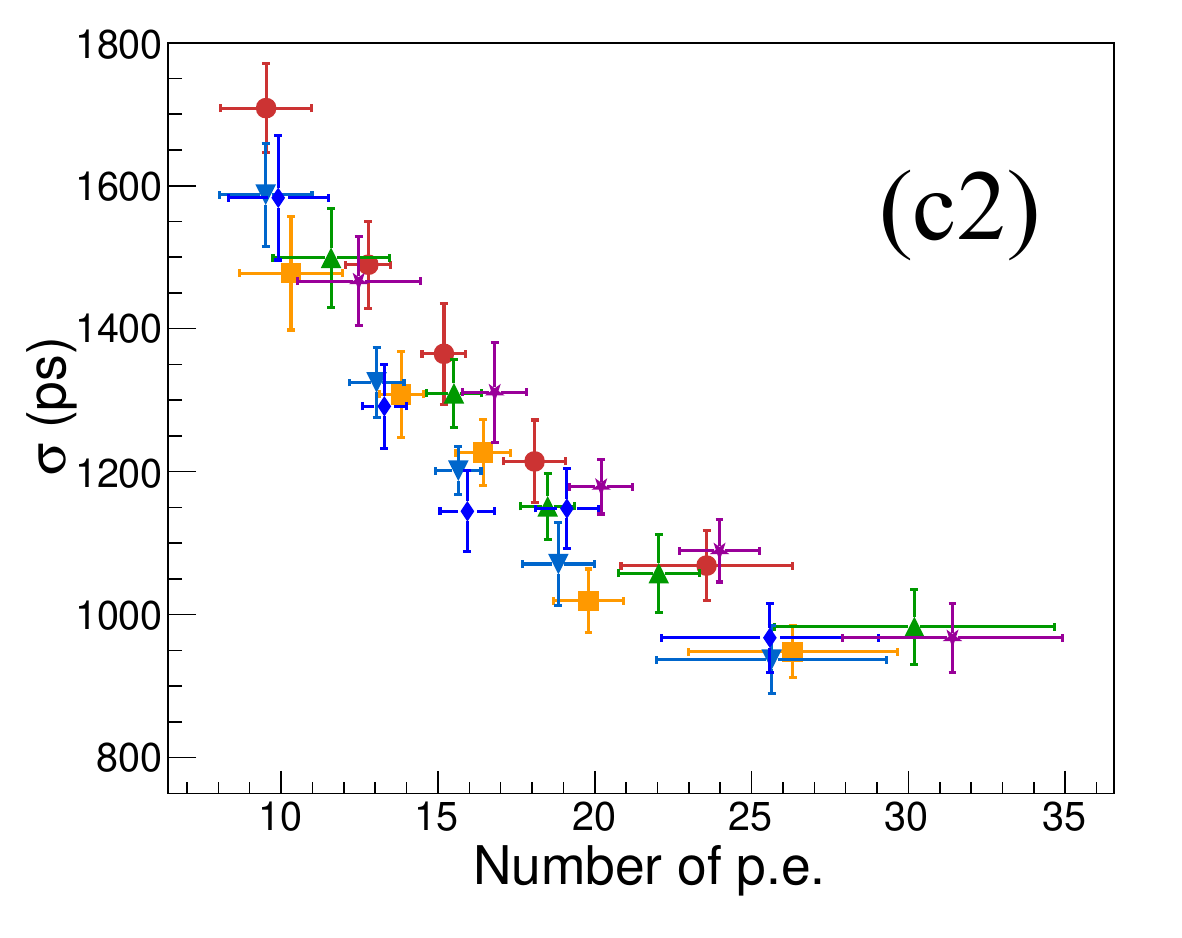}\\
\includegraphics[width=0.3\textwidth]{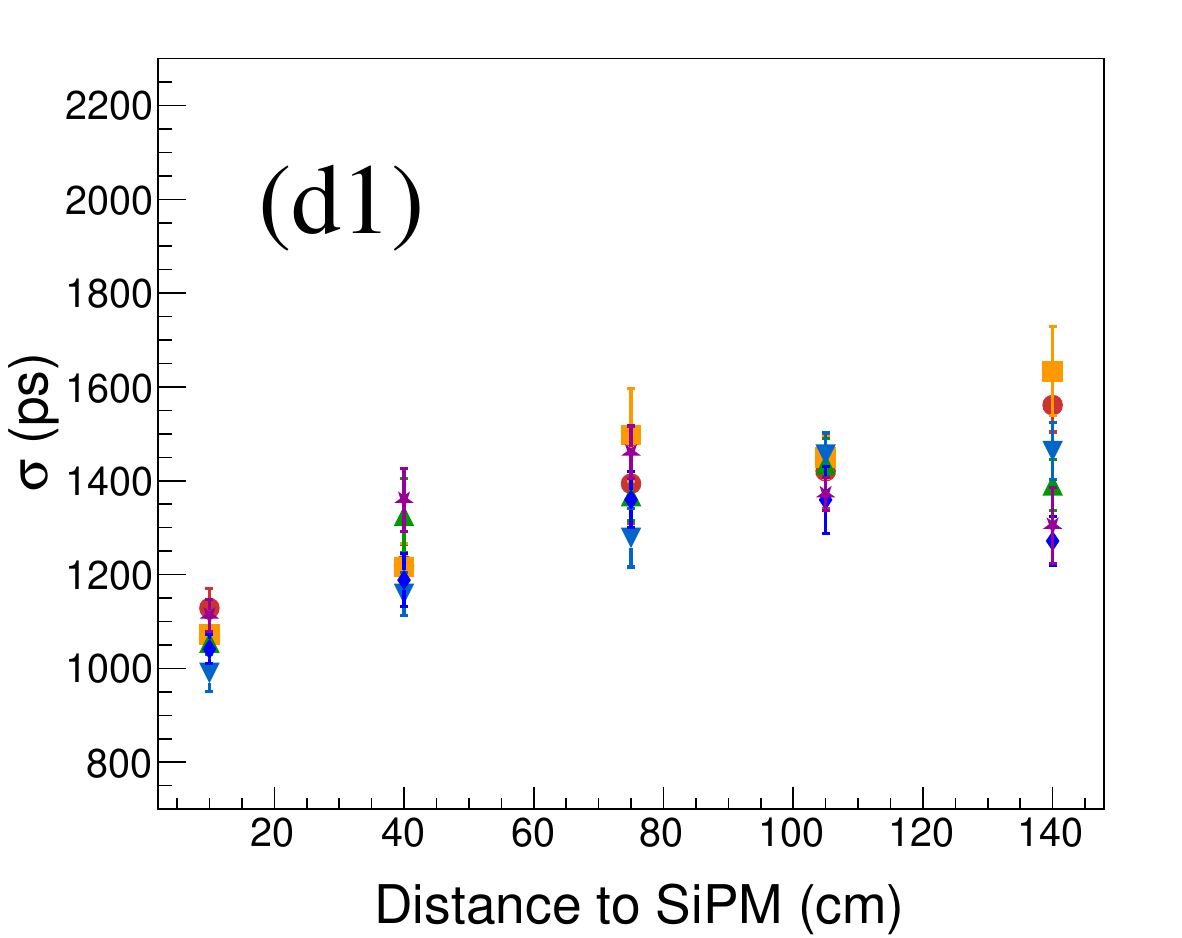} 
\includegraphics[width=0.3\textwidth]{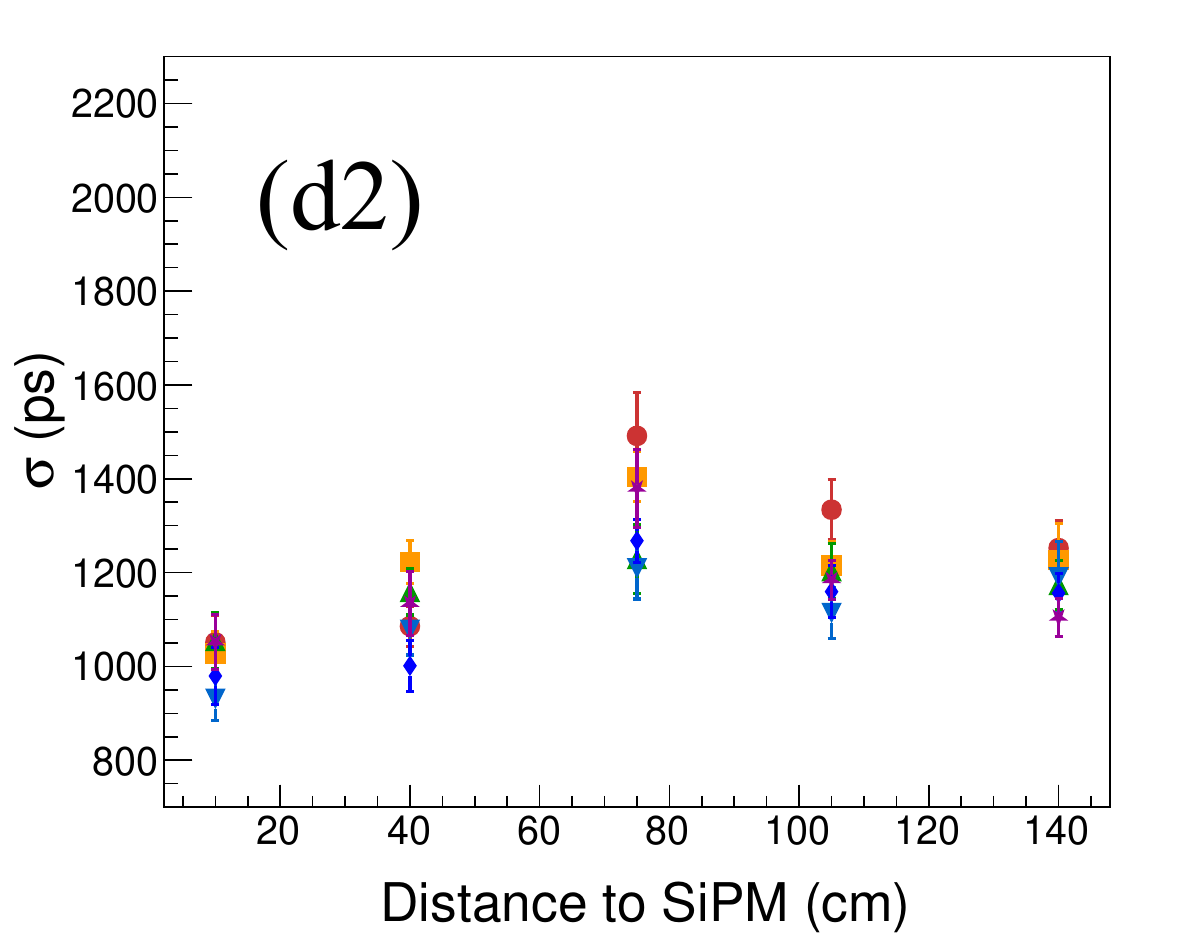} \\
\includegraphics[width=0.3\textwidth]{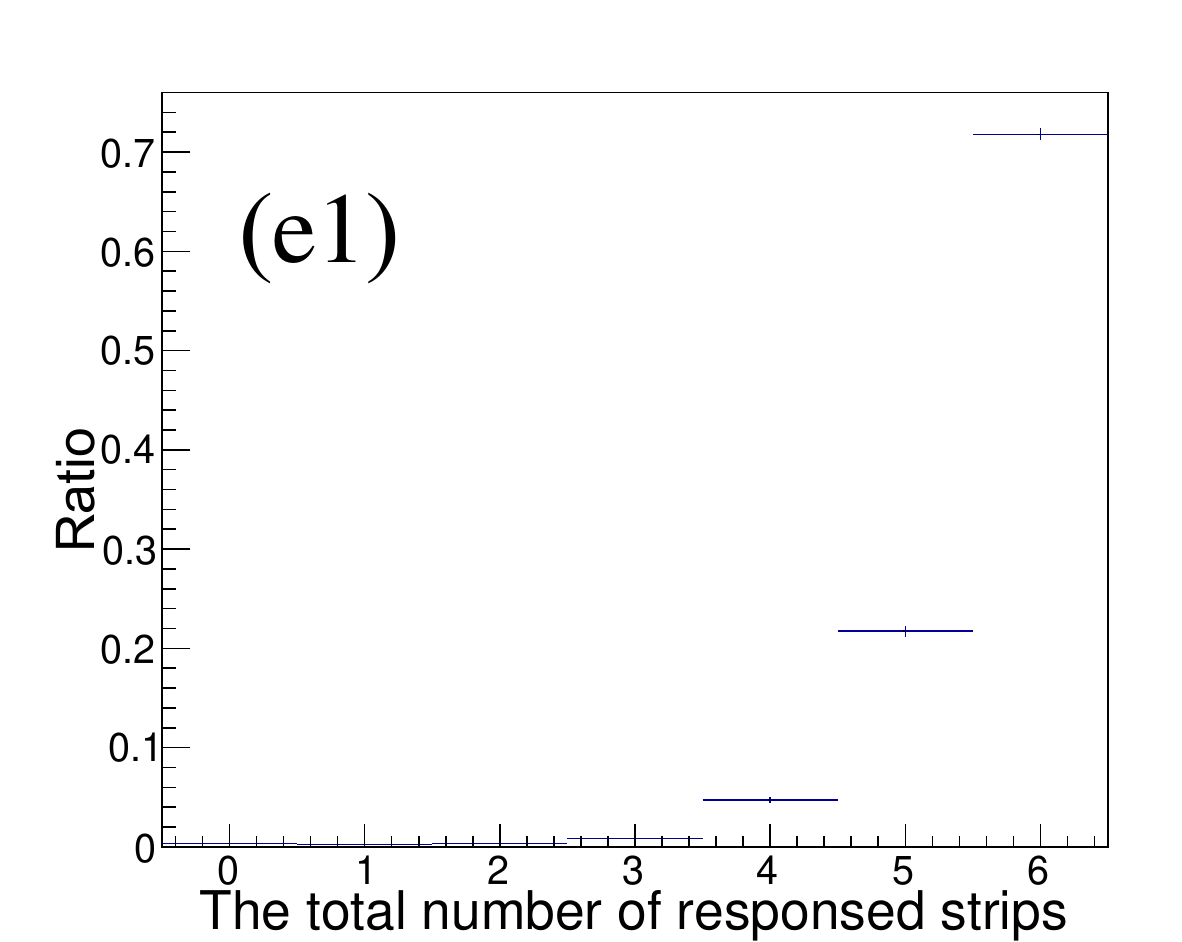} 
\includegraphics[width=0.3\textwidth]{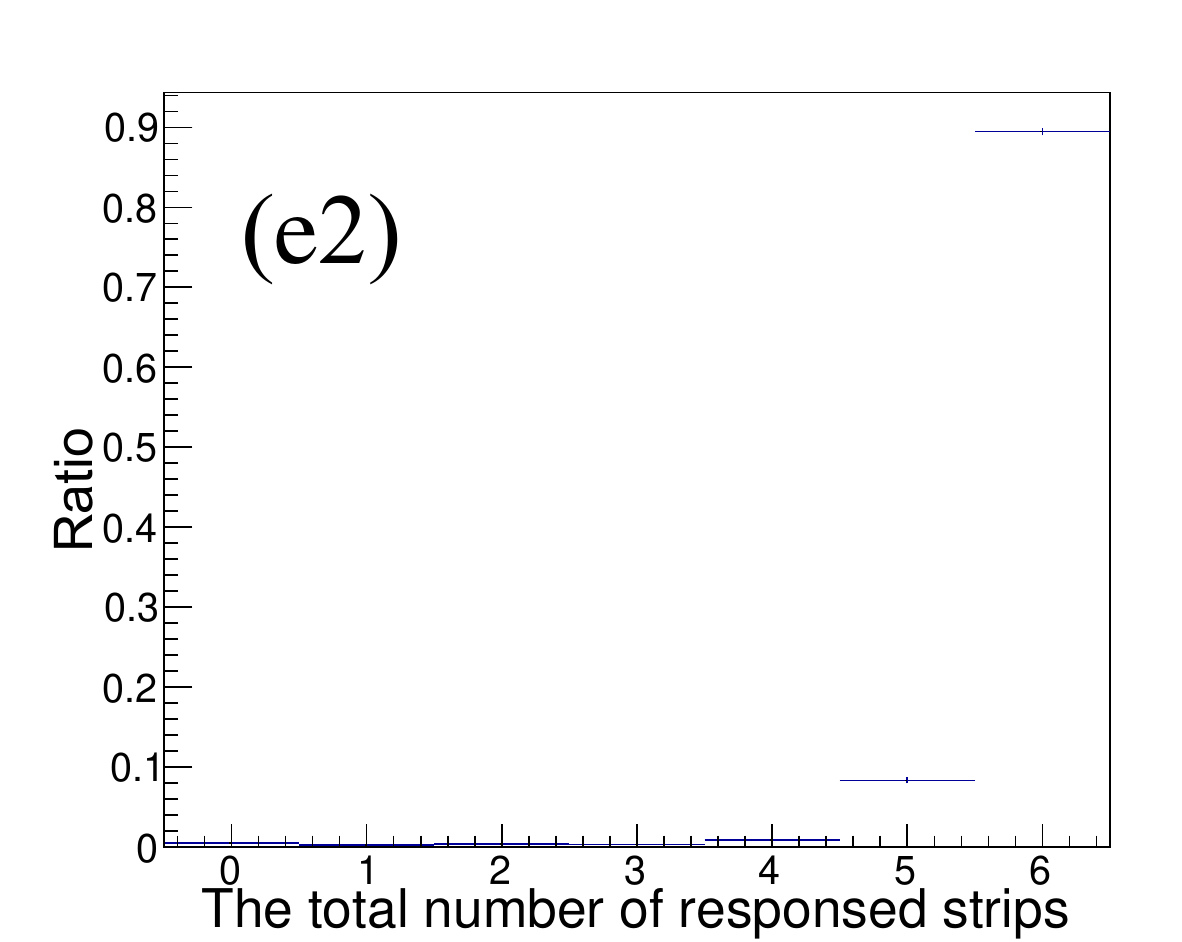} 
\ecntr
\caption{The performance of the detector arrays in cosmic ray tests is detailed, showcasing setups based on NDL 
SiPMs and Hamamatsu MPPCs, each with six detector channels.  Plots (a1) and (a2) are the distributions of $\npe$ from 
the cosmic ray hits along the strips, where the dots and the error bars are the means and the standard deviations 
from fitting to the $\npe$ distributions with Landau functions. Plots (b1) and (b2) are the efficiency versus the 
$\npe$ threshold at the far end. Plots (c1) and (c2) are the time resolutions versus the $\npe$ at far end. Plots 
(d1) and (d2) are the time resolutions triggering at the far ends of the strips. Plots (e1) and (e2) are the 
distributions of the number of detector channels responding to cosmic ray tracks. } 
\label{performances}
\end{figure}

Similar to the approach detailed in Fig.~\ref{comparison}, we evaluated the ADC distributions of the array at various 
positions and then converted these into distributions of $\npe$. We employed Landau functions for fitting the $\npe$ 
distributions and presented the resultant means and standard deviations in Figs.~\ref{performances}(a1) and 
\ref{performances}(a2). The array, equipped with NDL SiPMs, exhibited consistent photon collection performance, 
achieving up to 34 $\pe$s per cosmic ray event at the proximal end and 23 $\pe$s at the distal end, with the use of 
aluminum foil enhancing the reflective layer. Despite the variance in cross-talk levels, both NDL SiPMs and Hamamatsu 
MPPCs showed comparable photon collection efficiencies. The absence of a significant decline in $\npe$ beyond a 
$75~\cm$ distance underscores the advantage of the WLS fiber's extended attenuation length for maintaining high 
detection efficiency along the length of the scintillator strip.

As shown in Figs.~\ref{performances}(b1) and \ref{performances}(b2), efficiency assessments at the distant end for 
the six strips, fitted with either MPPCs or NDL SiPMs, revealed that most strips equipped with NDL SiPMs maintained 
efficiencies above 90\% at a threshold of $10.5~\pe$s. Strips with MPPCs also demonstrated high efficiency, exceeding 
84\% at the same threshold. These findings underscore the effectiveness of both NDL and MPPC SiPMs in capturing 
events at the far ends of the strips. It is important to note that some variance among the six strips can be 
attributed to the assembly quality of the SiPM and fiber.

Based on the results at operational voltages, the optical cross-talk for NDL EQR15 11-3030D-S SiPMs is around 20\%, 
while it is about 1\% for MPPC S13360. Performance evaluation using strip response thresholds shows that 90\% of 
cosmic rays trigger a response in all MPPC strips, compared to 70\% with NDL SiPMs. Achieving a cosmic ray detection 
efficiency of approximately 95\% with NDL SiPMs and 99\% with MPPCs indicates that the high-efficiency design 
criteria for CEPC are satisfied. This suggests the potential for a highly efficient, robust, and compact muon 
detector using scintillators, WLS fibers, and SiPMs.

A key advantage of a scintillator-based muon detector is its superior time resolution. The time resolution of KLM 
built with RPCs was not considered in Belle. With the time calibration, which started from the Belle II Fudan group,
the time resolution of Belle II KLM achieves about $4~\ns$ in endcaps and $8~\ns$ in the barrel, with ongoing work 
for further enhancements. Our study on detector channel time resolution used cosmic rays, employing a trigger system 
with long attenuation scintillator strips and four large SiPMs per strip, yielding about $60~\ps$ time resolution. By 
positioning triggers along the strip and measuring time differences, we calculated the strip's time resolutions. 
Results from MPPC S13360 and NDL EQR15 11-3030D-S tests are shown in Figs.~\ref{performances}(c1) and 
\ref{performances}(c2), demonstrating that time resolution at a strip's distal end strongly correlates with the 
$\npe$. Achieving a time resolution under $1~\ns$ requires $\npe > 35$ for NDL SiPMs and $\npe > 30$ for MPPCs, 
suggesting that enhancing the scintillator's light yield or using multiple fibers and large SiPMs can improve time 
resolution.

Figures~\ref{performances}(d1) and \ref{performances}(d2) illustrate the time resolutions across different positions 
on the long strips, with NDL SiPMs showing approximately $1.4~\ns$ and MPPCs about $1.2~\ns$. A linear fit to the 
time differences between signals from the long strip detector and the trigger, at various trigger positions, 
determined the scintillation photon propagation speed in Kuraray fiber as $(16.3\pm 2.8)~\cm/\ns$. 

\section{Summary and discussion}

For the CEPC muon detector's design, research on using extruded plastic scintillator, WLS fiber, and SiPM is 
conducted. The GNKD company's scintillators, Hamamatsu and NDL's SiPMs, and Kuraray's WLS fiber are used. 
SiPM performance enhancement and photon collection improvement are achieved through a coupling component for WLS 
fiber and SiPM attachment. Teflon is found to be effective as a reflection layer. The long scintillator strips 
demonstrate over 90\% efficiency and better than $1.7~\ns$ time resolution, indicating the technology's suitability 
for CEPC's muon detector construction. 

Based on the measurements described in this article, we are presently advancing the R\&D of the CEPC muon detector, 
aiming for the input for the RefTDR in the near future. This encompasses the construction of prototype modules with 
an area of about $4~\m^2$, the development of backend electronics for digitization, and the simulation of detector 
performance using Geant4. Enhancements in the detector design's performance are anticipated. The $3\mm \times 3\mm$ 
area ensures stable coupling between the SiPM and the WLS fiber. Plans include improving light yield by tuning the 
luminescent material, using NDL SiPMs with larger pixel size ($\ge 20~\um$), using Kurary WLS fiber with larger 
diameter ($2~\mm$) or using multiple fibers, with the expectation of improving photon collection, as well as 
enhancing detection efficiency and time resolution. They will help improve the performance of a muon detector based 
on extruded scintillator, WLS fiber, and SiPM, to increase the capabilities of muon identification, trigger based on 
muon tracks with time resolutions better than $1~\ns$, and searching for a long-lived particle using the large 
volume, the good spatial resolution, and the high time resolution of this detector. 

%\appendix
%\section{Some title}
%Please always give a title also for appendices.

\acknowledgments

This work is partially supported by the National Key R\&D Program of China under Contract No. 2022YFA1601903; 
National Natural Science Foundation of China under Contracts No. 11925502, No. 11961141003, and No. 12175041; and the 
Strategic Priority Research Program of the CAS under Contract No. XDB34030000. 

%\paragraph{Note added.} This is also a good position for notes added after the paper has been written.


\end{document}